\documentclass[12pt]{article}
\usepackage{amsmath,amsfonts,latexsym,amsthm,amssymb}
\newcommand{\DM}{\mathbb D^{(\mu )}}
\newcommand{\IM}{\mathbb I^{(\mu )}}
\newcommand{\K}{\mathcal K}
\newcommand{\LL}{\mathcal L}

\DeclareMathOperator{\I}{Im}
\DeclareMathOperator{\R}{Re}
\begin{document}
\newtheorem{teo}{Theorem}
\title{Distributed Order Calculus: an Operator-Theoretic Interpretation}
\author{Anatoly N. Kochubei
\footnote{Partially supported by the Ukrainian Foundation
for Fundamental Research, Grant 14.1/003.}
\\ \footnotesize Institute of Mathematics,\\
\footnotesize National Academy of Sciences of Ukraine,\\
\footnotesize Tereshchenkivska 3, Kiev, 01601 Ukraine}
\date{}
\maketitle

\vspace*{3cm}
\begin{abstract}
Within the functional calculi of Bochner-Phillips and Hirsch, we describe
the operators of distributed order differentiation and integration as functions
of the classical differentiation and integration operators respectively.
\end{abstract}

{\bf 1. Introduction and Preliminaries.}
In the distributed order calculus \cite{K}, used in physics for
modeling ultraslow diffusion and relaxation phenomena, we consider
derivatives and integrals of distributed order. The definitions
are as follows.

Let $\mu$ be a continuous non-negative function on $[0,1]$. The {\it
distributed order derivative} $\DM$ of weight $\mu$ for a function
$\varphi$ on $[0,T]$ is
\begin{equation}
\left( \DM \varphi \right) (t)=\int\limits_0^1 (\mathbb D^{(\alpha )}
\varphi )(t)\mu (\alpha )\,d\alpha
\end{equation}
where $D^{(\alpha )}$ is the Caputo-Dzhrbashyan regularized
fractional derivative of order $\alpha$, that is
\begin{equation}
\left( \mathbb D^{(\alpha )}\varphi \right) (t)=\frac{1}{\Gamma
(1-\alpha )}\left[ \frac{d}{dt}\int\limits_0^t(t-\tau )^{-\alpha
}\varphi (\tau )\,d\tau -t^{-\alpha }\varphi (0)\right],\quad
0<t\le T.
\end{equation}

Denote
\begin{equation}
k(s)=\int\limits_0^1\frac{s^{-\alpha }}{\Gamma (1-\alpha )}\mu
(\alpha )\,d\alpha ,\quad s>0.
\end{equation}
It is obvious that $k$ is a positive decreasing function. The
definition (1)-(2) can be rewritten as
\begin{equation}
\left( \DM \varphi\right) (t)=\frac{d}{dt}\int\limits_0^tk(t-\tau )
\varphi (\tau )\,d\tau -k(t)\varphi (0).
\end{equation}
The right-hand side of (4) makes sense for a continuous function
$\varphi$, for which the derivative $\dfrac{d}{dt}\int\limits_0^tk(t-\tau )
\varphi (\tau )\,d\tau$ exists.

If a function $\varphi$ is absolutely continuous, then
\begin{equation}
\left( \DM \varphi\right) (t)=\int\limits_0^tk(t-\tau )
\varphi' (\tau )\,d\tau .
\end{equation}

Below we always assume that $\mu \in C^3[0,1]$, $\mu (1)\ne 0$,
and either $\mu (0)\ne 0$, or $\mu (\alpha )\sim a\alpha^\nu $,
$a,\nu >0$, as $\alpha \to 0$. Under these assumptions (see
\cite{K}),
\begin{equation*}
k(s)\sim s^{-1}(\log s)^{-2}\mu (1),\quad s\to 0,
\end{equation*}
\begin{equation*}
k'(s)\sim -s^{-2}(\log s)^{-2}\mu (1),\quad s\to 0,
\end{equation*}
so that $k\in L_1(0,T)$ and $k$ does not belong to any $L_p$, $p>1$.
We cannot differentiate under the integral in (4), since $k'$ has a
non-integrable singularity.

It is instructive to give also the asymptotics of the Laplace transform
$$
\K (z)=\int\limits_0^\infty k(s)e^{-zs}ds.
$$
Using (4) we find that
$$
\K (z)=\int\limits_0^1z^{\alpha -1}\mu (\alpha )\,d\alpha,
$$
so that $\K (z)$ can be extended analytically to an analytic
function on $\mathbb C\setminus \mathbb R_-$, $\mathbb R_-=\{
z\in \mathbb C:\ \I z=0,\R z\le 0\}$. If $z\in \mathbb
C\setminus \mathbb R_-$, $|z|\to \infty$, then \cite{K}
\begin{equation}
\K (z)=\frac{\mu (1)}{\log z}+O\left( (\log |z|)^{-2}\right) ;
\end{equation}
see \cite{K} for further properties of $\K$.

The {\it distributed order integral} $\IM$ is defined as the
convolution operator
\begin{equation}
\left( \IM f\right) (t)=\int\limits_0^t\varkappa
(t-s)f(s)\,ds,\quad 0\le t\le T,
\end{equation}
where $\varkappa (t)$ is the inverse Laplace transform of the
function $z\mapsto \dfrac{1}{z\K (z)}$,
\begin{equation}
\varkappa (t)=\frac{d}{dt}\frac{1}{2\pi i}\int\limits_{\gamma
-i\infty }^{\gamma +i\infty }\frac{e^{zt}}{z}\cdot \frac{1}{z\K (z)}\,dz,
\quad \gamma >0.
\end{equation}
It was proved in \cite{K} that  $\varkappa \in C^\infty (0,\infty )$, and
$\varkappa$ is completely monotone; for small values of $t$,
\begin{equation}
\varkappa (t)\le C\log \frac{1}{t},\quad
|\varkappa' (t)|\le Ct^{-1}\log \frac{1}{t},
\end{equation}
If $f\in L_1(0,T)$, then $\DM \IM f=f$.

The aim of this paper is to clarify the operator-theoretic meaning
of the above constructions. It is well known that fractional
derivatives and integrals can be interpreted as fractional powers
of the differentiation and integration operators in various Banach
spaces; see, for example, \cite{BT,GK,JK,SKM}.

Let $A$ be the differential operator $Au=-\dfrac{du}{dx}$ in
$L_p(0,T)$, $1\le p<\infty$, with the boundary condition $u(0)=0$.
Its domain $D(A)$ consists of absolutely continuous functions
$u\in L_p(0,T)$ , such that $u(0)=0$ and $u'\in L_p(0,T)$. We show
that on $D(A)$ the distributed order differentiation coincides
with the function $\LL (-A)$ of the operator $-A$, where $\LL
(z)=z\K (z)$, and the function of an operator is understood in the
sense of the Bochner-Phillips functional calculus (see
\cite{Ph,BBD,Sch}).

Moreover, if $p=2$, then the distributed order integration
operator $\IM$ equals $\mathcal N(J)$, where $\mathcal
N(x)=\dfrac{1}{\LL (x)}$, $J$ is the integration operator,
$(Ju)(t)=\int\limits_0^t u(\tau )\,d\tau$. This result is obtained
within Hirsch's functional calculus \cite{H1,H2} giving more
detailed results for a more narrow class of functions. As
by-products, we obtain an estimate of the semigroup generated by
$-\LL (-A)$, and an expression for the resolvent of the operator
$\IM$.

\bigskip
{\bf 2. Functions of the differentiation operator.} The semigroup
$U_t$ of operators on the Banach space $X=L_p(0,T)$ generated by
the operator $A$ has the form
$$
\left( U_tf\right) (x)=\begin{cases}
f(x-t), & \text{ if $0\le t\le x<T$;}\\
0, & \text{ if $0<x<t$,}\end{cases}
$$
$x\in (0,T)$, $t\ge 0$. This follows from the easily verified
formula for the resolvent $R(\lambda ,A)=(A-\lambda I)^{-1}$ of
the operator $A$:
\begin{equation}
(R(\lambda ,A)u)(x)=-\int\limits_0^xe^{-\lambda (x-y)}u(y)\,dy;
\end{equation}
see \cite{Kato} for a similar reasoning for operators on
$L_p(0,\infty )$. The semigroup $U_t$ is nilpotent, $U_t=0$ for
$t>T$; compare Sect. 19.4 in \cite{HP}. It follows from the
expression (10) and the Young inequality that $\| R(\lambda ,A)\|
\le \lambda^{-1}$, $\lambda >0$, so that $U_t$ is a $C_0$-semigroup
of contractions.

In the Bochner-Phillips functional calculus, for the operator $A$,
as a generator of a contraction semigroup, and any function $f$ of
the form
\begin{equation}
f(x)=\int\limits_0^\infty \left( 1-e^{-tx}\right) \sigma
(dt)+a+bx,\quad a,b\ge 0,
\end{equation}
where $\sigma$ is a measure on $(0,\infty )$, such that
$$
\int\limits_0^\infty \frac{t}{1+t}\sigma (dt)<\infty ,
$$
the subordinate $C_0$-semigroup $U_t^f$ is defined by the Bochner
integral
$$
U_t^f=\int\limits_0^\infty \left( U_s u\right) \sigma_t(ds)
$$
where the measures $\sigma_t$ are defined by their Laplace
transforms,
$$
\int\limits_0^\infty e^{-sx}\sigma_t (ds)=e^{-tf(x)}.
$$

The class $\mathcal B$ of functions (11) coincides with the class
of Bernstein functions, that is functions $f\in C([0,\infty )\cap
C^\infty (0,\infty )$, for which $f'$ is completely monotone.
Below we show that $\LL \in \mathcal B$.

The generator $A^f$ of the semigroup $U_t^f$ is identified with
$-f(-A)$. On the domain $D(A)$,
\begin{equation}
A^fu=-au+bAu+\int\limits_0^\infty (U_tu-u)\sigma (dt),\quad u \in D(A).
\end{equation}

\medskip
\begin{teo}
$\mathrm{(i)}$ If $u\in D(A)$, then $A^\LL u=-\DM u$.

$\mathrm{(ii)}$ The semigroup $U_t^\LL$ decays at infinity faster
than any exponential function:
\begin{equation}
\left\| U_t^\LL \right\| \le C_re^{-rt}\quad \text{for any $r>0$}.
\end{equation}
The operator $A^\LL$ has no spectrum.

$\mathrm{(iii)}$ The resolvent $R(\lambda ,-A^\LL )$ of the
operator $-A^\LL$ has the form
\begin{equation}
\left( R(\lambda ,-A^\LL )u\right) (x)=\int\limits_0^xr_\lambda
(x-s)u(s)\, ds,\quad u\in X,
\end{equation}
where
\begin{equation}
r_\lambda (s)=\frac{1}\lambda \frac{d}{ds}u_\lambda (s),
\end{equation}
and $u_\lambda$ is the solution of the Cauchy problem
\begin{equation}
\DM u_\lambda =\lambda u_\lambda ,\quad u_\lambda (0)=1.
\end{equation}

$\mathrm{(iv)}$ The inverse $\left( -A^\LL \right)^{-1}$ coincides
with the distributed order integration operator $\IM$.

$\mathrm{(v)}$ The resolvent of $\IM$ has the form
\begin{equation}
(\IM -\lambda I)^{-1}u=-\frac{1}\lambda
u-\frac{1}{\lambda^2}r_{1/\lambda}*u,\quad \lambda \ne 0.
\end{equation}
\end{teo}

\medskip
{\it Proof}. Let $\sigma (dt)=-k'(t)\,dt$. By (3),
$$
k'(t)=-\int\limits_0^1\frac{\alpha t^{-\alpha -1}}{\Gamma
(1-\alpha )}\mu (\alpha )\,d\alpha ,
$$
so that
$$
\int\limits_0^\infty \frac{t}{1+t}\sigma (dt)=\int\limits_0^1
\frac{\alpha \mu (\alpha )}{\Gamma
(1-\alpha )}\,d\alpha \int\limits_0^\infty \frac{t^{-\alpha
}}{1+t}\,dt.
$$
Using the integral formula 2.2.5.25 from \cite{PBM} we find that
$$
\int\limits_0^\infty \frac{t}{1+t}\sigma (dt)=\pi \int\limits_0^1
\frac{\alpha \mu (\alpha )}{(\sin \alpha \pi )\Gamma
(1-\alpha )}\,d\alpha <\infty .
$$

Let us compute the function (11) with $a=b=0$. We have
$$
f(x)=-\int\limits_0^\infty \left( 1-e^{-tx}\right) k'(t)\,dt =x\int\limits_0^\infty
e^{-tx}k(t)\,dt=x\K (x)=\LL (x).
$$
The corresponding expression (12) for $A^\LL u$, $u\in D(A)$, is
as follows:
\begin{multline*}
\left( A^\LL u\right) (x)=-\int\limits_0^\infty
[(U_tu)(x)-u(x)]k'(t)\,dt=-\int\limits_0^x[u(x-t)-u(x)]k'(t)\,dt
+u(x)\int\limits_x^\infty k'(t)\,dt\\
=-k(x)u(x)-\int\limits_0^x[u(x-t)-u(x)]k'(t)\,dt.
\end{multline*}
By (4), we find that $A^\LL u=-\DM u$, $u\in D(A)$.

The function $\LL (z)$ is holomorphic for $\R z>0$. We will need a
detailed information (refining (6)) on the behavior of $\R \LL
(\sigma +i\tau )$, $\sigma ,\tau \in \mathbb R$, $\sigma >0$, when
$|\tau |\to \infty$. We have
$$
\R \LL (\sigma +i\tau ) =\int\limits_0^1\varphi (\alpha ,\sigma
,\tau )\mu (\alpha )\,d\alpha
$$
where
$$
\varphi (\alpha ,\sigma ,\tau )=(\sigma^2+\tau^2)^{\alpha /2}\cos
\left( \alpha \arctan \frac{\tau }\sigma \right).
$$

We check directly that $\varphi (\alpha ,\sigma
,0)=\sigma^\alpha$,
$$
\frac{\partial \varphi (\alpha ,\sigma ,\tau )}{\partial \tau }
=\alpha (\sigma^2+\tau^2)^{\alpha /2-1}\cos \left( \alpha \arctan
\frac{\tau }\sigma \right) \left[ \tau -\sigma \tan \left( \alpha \arctan \frac{\tau
}\sigma \right) \right]\ge 0,
$$
and $\dfrac{\partial \varphi (\alpha ,\sigma ,\tau )}{\partial \tau
} >0$ for $\alpha <1$. This means that the function $g_\sigma
(\tau )=\R \LL (\sigma +i\tau )$ (which is even in $\tau$) is
strictly monotone increasing in $\tau$ for $\tau >0$. Its minimal
value is
$$
g_\sigma (0)=\int\limits_0^1\sigma^\alpha \mu (\alpha )\,d\alpha .
$$
On the other hand,
\begin{multline*}
\R \LL (\sigma +i\tau )\ge \int\limits_0^1 (\sigma^2+\tau^2)^{\alpha /2}\cos
\frac{\alpha \pi }2\mu (\alpha )\,d\alpha =\frac{2}\pi
\int\limits_0^{\pi /2}(\sigma^2+\tau^2)^{t/\pi }\mu \left(
\frac{2t}{\pi }\right) \cos t \,dt \\
=\frac{2}\pi \int\limits_0^{\pi /2}e^{qt}\mu \left(
\frac{2t}{\pi }\right)\cos t  \,dt=\frac{2}\pi e^{q\pi /2}\int\limits_0^{\pi /2}
e^{-qs}\mu \left( 1-\frac{2}\pi s\right) \sin s\,ds
\end{multline*}
where $q=\frac{1}\pi \log (\sigma^2+\tau^2)$. By Watson's
asymptotic lemma (see \cite{Olver}), since $\mu (1)\ne 0$, we have
$$
\int\limits_0^{\pi /2}
e^{-qs}\mu \left( 1-\frac{2}\pi s\right)\sin s \,ds\sim Cq^{-2}
$$
where $C$ does not depend on $\sigma ,\tau$. Roughening the
estimate a little we find that
\begin{equation}
e^{-t\R \LL (\sigma +i\tau )}\le Ce^{-t\rho |\tau
|^{\frac{1}2-\varepsilon }}
\end{equation}
where $0<\varepsilon <\frac12$ can be taken arbitrarily, and the
positive constants $C$ and $\rho$ do not depend on $\sigma$ and
$\tau$.

It follows from (18) (see \cite{DP}) that for each $t>0$ the
function $x\mapsto e^{-t\LL (x)}$ is represented by an absolutely
convergent Laplace integral. This means that the measure
$\sigma_t(ds)$ has a density $m(t,s)$ with respect to the Lebesgue
measure. Moreover,
\begin{equation}
m(t,s)=\frac{1}{2\pi i}\int\limits_{\gamma -i\infty }^{\gamma +i\infty }
e^{zs}\cdot e^{-t\LL (z)}\,dz,\quad \gamma >0.
\end{equation}
Since $U_t=0$ for $t>T$, we have
\begin{equation}
U_t^fu=\int\limits_0^T(U_su)m(t,s)\,ds.
\end{equation}

The representation (19) yields the expression
$$
m(t,s)=\frac{e^{\gamma s}}\pi \int\limits_0^\infty e^{i\tau }
e^{-t\LL (\gamma +i\tau )}\,d\tau,
$$
$$
\LL (\gamma +i\tau )=\int\limits_0^1(\gamma +i\tau )^\alpha \mu
(\alpha )\,d\alpha ,\quad 0\le \tau <\infty .
$$
We have
$$
|m(t,s)|\le \frac{e^{\gamma s}}\pi \int\limits_0^\infty
e^{-tg_\gamma (\tau )}\,d\tau .
$$

The above monotonicity property of $g_\gamma$ makes it possible to
apply to the last integral the Laplace asymptotic method
\cite{Olver}. We obtain that, for large values of $t$,
$$
|m(t,s)|\le Ct^{-1}e^{\gamma s}e^{-tg_\gamma (0)}.
$$
Changing $\gamma$ and $C$ we can make the coefficient $g_\gamma
(0)$ arbitrarily big. By (20), this leads to the estimate (13).

Due to (13), the resolvent
\begin{equation}
R(\lambda ,A^\LL )=-\int\limits_0^\infty e^{-\lambda t}U_t^\LL \,dt,
\end{equation}
is an entire function, so that $A^\LL$ has no spectrum.

It follows from (21) that
$$
R(\lambda ,-A^\LL )=\int\limits_0^\infty e^{\lambda t}U_t^\LL \,dt,
$$
and if $u\in X$, $\R \lambda \le 0$, then
$$
\left( R(\lambda ,-A^\LL )u\right) (x)=\int\limits_0^\infty e^{\lambda t}
\,dt\int\limits_0^xu(x-s)m(t,s)\,ds=\int\limits_0^xr_\lambda (x-s)u(s)\,ds
$$
where
\begin{equation}
r_\lambda (s)=\int\limits_0^\infty e^{\lambda t}m(t,s)\,dt.
\end{equation}

For a fixed $\omega \in (\frac12,1)$, let us deform the contour of
integration in (19) from the vertical line to the contour
$S_{\gamma ,\omega}$ consisting of the arc
$$
T_{\gamma ,\omega }=\{ z\in \mathbb C:\ |z|=\gamma ,|\arg z|\le
\omega \pi \},
$$
and two rays
$$
\Gamma_{\gamma ,\omega }^\pm =\{ z\in \mathbb C:\ |\arg z|=\pm
\omega \pi ,|z|\ge \gamma \}.
$$
The contour $S_{\gamma ,\omega}$ is oriented in the direction of
growth of $\arg z$. By Jordan's lemma,
$$
m(t,s)=\frac{1}{2\pi i}\int\limits_{S_{\gamma ,\omega}}
e^{zs}\cdot e^{-t\LL (z)}\,dz.
$$
Under this integral, we may integrate in $t$, as required in (22).
We find that
\begin{equation}
r_\lambda (s)=\frac{1}{2\pi i}\int\limits_{S_{\gamma ,\omega}}
\frac{e^{zs}}{\LL (z)-\lambda }\,dz,\quad s>0
\end{equation}
(for $\R \lambda >0$, $\gamma$ should be taken big enough).

If $\lambda =0$, the right-hand side of (23) coincides with that of (8)
(see also the formula (3.4) in \cite{K}), and we prove that
$\left( -A^\LL \right)^{-1}=\IM$.

For $\lambda \ne 0$, we rewrite (23) as
\begin{equation}
r_\lambda (s)=\frac{1}{2\pi i\lambda }\int\limits_{S_{\gamma ,\omega}}
e^{zs}\frac{\LL (z)}{\LL (z)-\lambda }\,dz-
\frac{1}{2\pi i\lambda }\int\limits_{S_{\gamma ,\omega}}
e^{zs}\,dz.
\end{equation}
For $0<s<T$, we have
$$
\int\limits_{S_{\gamma ,\omega}}e^{zs}\,dz=-\lim\limits_{R\to
\infty }\int\limits_{\substack{|z|=R\\ \omega \pi <|\arg z|<\pi}}
e^{zs}\,dz,
$$
$$
\left| \int\limits_{\substack{|z|=R\\ \omega \pi <|\arg z|<\pi}}
e^{zs}\,dz\right| \le 2R\int\limits_{\omega \pi }^\pi e^{Rs\cos
\varphi }\,d\varphi \le 2R\pi (1-\omega )e^{Rs\cos \omega \pi}\to
0,
$$
as $R\to \infty$.

Thus, the second integral in (24) equals zero, and it remains to
compare (24) with the formula (2.15) of \cite{K} giving an
integral representation of the function $u_\lambda$.

The formula (17) follows from (15) and the general connection
between the resolvents of an operator and its inverse
(\cite{Kato}, Chapter 3, formula (6.18)). The theorem is proved.

\medskip
Note that the expression (17) for the resolvent of a distributed
order integration operator is quite similar to the Hille-Tamarkin
formula for the resolvent of a fractional integration operator
(see \cite{HP}, Sect. 23.16). In our case, the function
$u_\lambda$ is a counterpart of the function $z\mapsto E_\alpha
(\lambda z^\alpha )$ (for the order $\alpha$ case). However, in our
situation no analog of the entire function $E_\alpha$ (the Mittag-Leffler
function) has been identified so far. Accordingly, our proof of (17)
is different from the reasoning in \cite{HP}.

\bigskip
{\bf 3. Functions of the integration operator.} In this section we assume
that $p=2$.

Hirsch's functional calculus deals with the class $\mathcal R$ of
functions which are continuous on $\mathbb C\setminus (-\infty
,0)$, holomorphic on $\mathbb C\setminus (-\infty ,0]$, transform
the upper half-plane into itself, and transform the semi-axis
$(0,\infty )$ into itself. The class $\mathcal R$ is a subclass of
$\mathcal B$.

Another important class of functions is the class $\mathcal S$ of
Stieltjes functions
$$
f(z)=a+\int\limits_0^\infty \frac{d\rho (\lambda )}{z+\lambda
},\quad z\in \mathbb C\setminus (-\infty ,0],
$$
where $a\ge 0$, $\rho$ is a non-decreasing right-continuous
function, such that $\int\limits_0^\infty\dfrac{d\rho
(t)}{1+t}<\infty$. If $f$ is a nonzero function from $\mathcal S$,
then the function
$$
\widetilde{f}(z)=\frac{1}{f(z^{-1})}
$$
also belongs to $\mathcal S$.

If $f\in \mathcal S$, then the function $H_f(z)=f(z^{-1})$ belongs
to $\mathcal R$. It has the form
$$
H_f(z)=a+\int\limits_0^\infty \frac{z}{1+\lambda z}d\rho (\lambda ),
\quad z\in \mathbb C\setminus (-\infty ,0].
$$

For some classes of linear operators $V$, the function $H_f(V)$ is
defined as a closure of the operator
$$
Wx=ax+\int\limits_0^\infty V(I+\lambda V)^{-1}xd\rho (\lambda
),\quad x\in D(V).
$$
In particular, this definition makes sense if $-V$ is a generator
of a contraction $C_0$-semigroup, and in this case the above
construction is equivalent to the Bochner-Phillips functional
calculus \cite{BBD,GK}. In addition, by Theorem 2 of \cite{H1}, if
$(-V)^{-1}$ is also a generator of a contraction $C_0$-semigroup,
then
\begin{equation}
\left[ H_f(V)\right]^{-1}=H_{\widetilde{f}}\left( V^{-1}\right) .
\end{equation}

In order to apply the above theory to our situation, note that \cite{H1}
$$
z^\alpha =\frac{1}{\Gamma (\alpha )\Gamma (1-\alpha )}\int\limits_0^\infty
\frac{z}{1+\lambda z}\lambda^{-\alpha }\,d\lambda ,\quad 0<\alpha <1,
$$
whence
$$
\LL (z)=\int\limits_0^\infty \frac{z}{1+\lambda z}\beta (\lambda )\,d\lambda
$$
where
$$
\beta (\lambda )=\int\limits_0^1\frac{\lambda^{-\alpha }\mu
(\alpha )}{\Gamma (\alpha )\Gamma (1-\alpha )}\,d\alpha .
$$
Thus $\LL (z)=H_f(z)$, with
$$
f(z)=\int\limits_0^\infty \frac{1}{z+\lambda }\beta (\lambda
)\,d\lambda.
$$
It follows from Watson's lemma \cite{Olver} that $\beta (\lambda
)\le C(\log \lambda )^{-2}$ for large values of $\lambda$.
Therefore
$$
\int\limits_0^\infty\frac{\beta (\lambda )}{1+\lambda }\,d\lambda
<\infty .
$$
Denote $\mathcal N(z)=H_{\widetilde{f}}(z)=\dfrac{1}{\LL (z)}$.

If $V=-A$, then $(-V)^{-1}=-J$, where $J$ is the integration
operator. It is easy to check that $\langle (J+J^*)u,u\rangle \ge
0$ ($\langle \cdot ,\cdot \rangle$ is the inner product in
$L_2(0,T)$). Therefore $-J$ is a generator of a contraction
semigroup.

After these preparations, the equality (25) implies the following
result.

\begin{teo}
The operator $\IM$ of distributed order integration and the
integration operator $J$ are connected by the relation
$$
\IM =\mathcal N(J).
$$
\end{teo}

\end{document}